\documentclass[journal=apchd5,manuscript=article]{achemso}
\usepackage{chemformula}
\usepackage[T1]{fontenc} \usepackage{graphicx,ulem,psfrag,amsmath,amssymb,amsfonts,latexsym,color,dcolumn,bm}
\usepackage[hidelinks]{hyperref}
\author{Wilton J. M. Kort-Kamp}
\email{kortkamp@lanl.gov}
\affiliation{Theoretical Division and Center for Nonlinear Studies, Los Alamos National Laboratory, MS B262, Los Alamos, New Mexico 87545, USA}
\author{Shobhita Kramadhati}
\affiliation{Center for Integrated Nanotechnologies, Los Alamos National Laboratory, MS K771, Los Alamos, New Mexico 87545, USA}
\author{Abul K. Azad}
\affiliation{Center for Integrated Nanotechnologies, Los Alamos National Laboratory, MS K771, Los Alamos, New Mexico 87545, USA}
\author{Matthew T. Reiten}
\affiliation{Los Alamos National Laboratory, P.O. Box 1663, New Mexico 87545, USA}
\author{Diego A. R. Dalvit}
\affiliation{Theoretical Division, Los Alamos National Laboratory, MS B213,  Los Alamos, New Mexico 87545, USA}
\title{Passive radiative "thermostat" enabled by phase-change photonic nanostructures}
\begin{document}
\begin{abstract}
A thermostat senses the temperature of a physical system and switches heating or cooling devices on or off, regulating the flow of heat to maintain the system's temperature near a desired
setpoint. Taking advantage of recent advances in radiative heat transfer technologies, here we propose a passive radiative "thermostat" based on phase-change photonic nanostructures for thermal regulation at room temperature. By self-adjusting their visible to mid-IR absorptivity and emissivity responses depending on the ambient temperature, the proposed devices use the sky to passively cool or heat during day-time using the phase-change transition temperature as the setpoint, while at night-time temperature is maintained at or below ambient. We simulate the performance of a passive nanophotonic thermostat design based on vanadium dioxide thin films, showing daytime passive cooling (heating) with respect to ambient in hot (cold) days, maintaining an equilibrium temperature approximately locked within the phase transition region. Passive radiative thermostats can potentially enable novel thermal management technologies, e.g. to moderate diurnal temperature in regions with extreme annual thermal swings.
\end{abstract}
\pagebreak

Passive radiative cooling uses the atmosphere's transparency windows in the mid-IR to channel heat into outer space, and can serve as an efficient technique to radiatively cool building structures 
\cite{Trombe1967,Catalanotti1975,Bartoli1977,Granqvist1981,Gentle2010,Zhu2015}, renewable energy harvesting devices \cite{Zhu2014,Safi2015}, and even textiles \cite{Hsu2016}. 
Since ancient times it is known that a black radiator facing a clear night sky could cool below hot ambient temperatures \cite{Bahadori1978}. However, daytime passive radiative cooling below ambient temperature of a surface under direct sunlight  poses much more stringent requirements due to heat absorption of solar radiation, and only recently high perfomance photonic structures that accomplish this goal have been theoretically proposed \cite{Rephaeli2013,Hossain2015} and experimentally realized \cite{Raman2014,Chen2016,Zhai2017}. These include nanostructured multilayered coolers that simultaneously act as optimized solar reflectors and mid-infrared (mid-IR) thermal emitters \cite{Raman2014}, as well as randomized glass-polymer hybrid films that are  transparent to the solar spectrum and also behave as good emitters across the atmospheric IR transparency windows \cite{Zhai2017}. On the other hand, high solar absorption is sometimes a desirable feature, particularly in energy harvesting applications such as solar thermophotovoltaics \cite{Harder2003,Rephaeli2009}. Various nanophotonic broadband solar absorbers have also been recently demonstrated based on metamaterials \cite{Hedayati2011,Azad2016}, dense nanorods and nanotube films \cite{Xi2007,Lenert2014}, multilayer planar photonic structures \cite{Shimizu2015, Deng2015}, and refractory plasmonics \cite{Li2014}. 

Although both passive radiative coolers and heaters offer clear advantages in the above technologies, they may also impose severe limitations to temperature management applications, e.g. maintaining buildings at temperatures moderate enough for human occupancy or to minimize stress on structures due to thermal cycling. For example, a typical passive radiative cooler will still cool during daytime even in cold days, while a standard passive radiative heater will heat to undesirable high temperatures in hot days. In this paper, we propose a nanophotonic passive radiative thermostat that effectively monitors the ambient temperature and self-adjusts its emissivity and absorptivity across the visible to mid-IR wavelength range to passively cool or heat in order to moderate the equilibrium temperature of a structure. To achieve this goal the device should reject solar radiation while increasing its mid-IR emissivity for  temperatures above a pre-set threshold, hence operating as a cooler. On the other hand, for temperatures below the threshold  the device should behave as a heater by absorbing sun light and minimizing heat losses via the atmospheric transparency windows. Thermochromic phase change materials are potential candidates for realizing the required absorption and emission properties. Previous works in the literature  have touched upon the concept of passive smart infrared emitters based on thermochromic materials for temperature regulation (see, for instance, \cite{Benkahoul2011,Kats2013, Biehs2015} and references therein). However, a detailed description of a passive radiative thermostat operating in the visible to infrared frequency range, allowing for cooling or heating with respect to ambient temperature, remains unexplored.

A commonly investigated phase change material is vanadium dioxide (VO$_2$), which is a correlated electron transition metal oxide that exhibits a reversible first-order insulator-to-metal phase transition at  $T_{\rm PC}^{\rm VO_2} \simeq 68 ^o$C for bulk samples\cite{Morin1959, Qazilbash2007}. For temperatures below the phase transition, VO$_2$ behaves as a narrow-gap semiconductor and it is transparent at IR wavelengths, while for temperatures above, it is highly reflective due to its metallic character. This switchable radiative response of VO$_2$ has been employed in the past for manipulating thermal radiation in the near- and far-field regime, including the development of thermal diodes \cite{Biehs2013, Ito2014, Fiorino2018}, transistors \cite{Biehs2014}, and memories \cite{Kubytskyi2014,Ito2016}. Temperature-modulated near-IR transmittance coatings for smart windows \cite{Gao2012, Powell2016, Smith2016} and tunable IR thermal emitters \cite{Wang2014, Taylor2017} based on VO$_2$ have also been reported. However, cooling below ambient temperature is usually not possible in these systems due to their inability to reject incident visible sun light. Another disadvantage of bulk vanadium dioxide is its high transition temperature, that makes it unpractical for thermal management at usual ambient conditions. Recent experimental works have demonstrated that it is possible to tune $T_{\rm PC}^{\rm VO_2}$ over a wide range of temperatures employing different approaches, including strain \cite{Wei2009, Cao2009}, pulsed voltage \cite{Driscoll2009},  doping \cite{Tan2012,Lee2013}, and nanostructuring \cite{Lopez2002, Lopez2004,Whittaker2009, Yoon2016, Muraoka2002, Fan2014, Liu2015}. Particularly, phase transition temperatures as low as $T_{\rm PC}^{\rm VO_2} \simeq 17 ^o$C have been reported in oriented VO$_2$ thin films on TiO$_2$ substrates \cite{Muraoka2002, Fan2014, Liu2015}. Such high degree of control over $T_{\rm PC}^{\rm VO_2}$ makes vanadium dioxide a promising material to achieve passive temperature management under a variety of climate conditions.

\begin{figure}
\centering
\includegraphics[width=8.6cm]{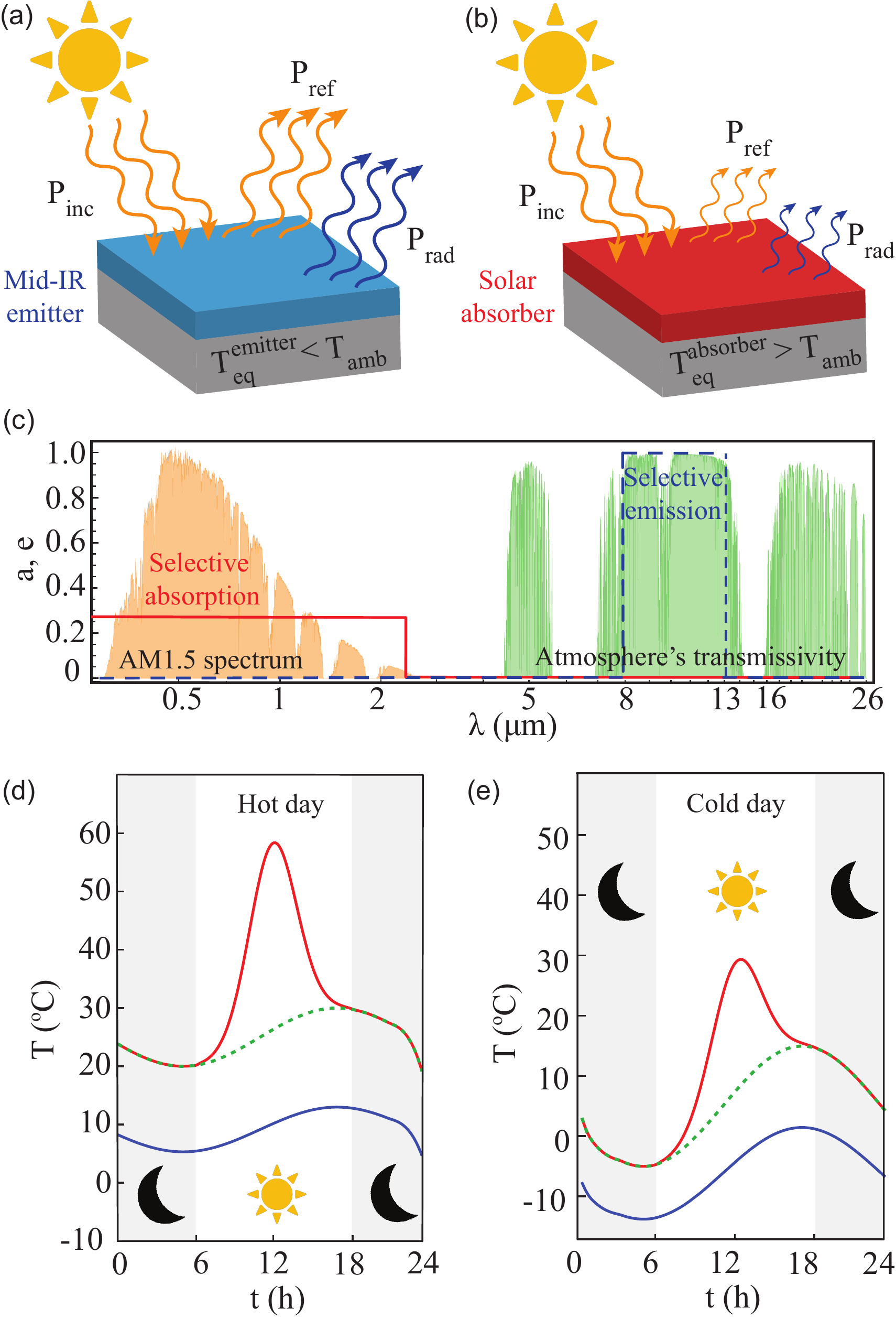}
\caption{Temperature management with passive photonic devices. Schematic representation of a  mid-infrared emitter (a) and a solar absorber (b) with idealized selective emissivity (blue) and absorptivity (red), as show in (c). The orange and green backgrounds correspond to the AM1.5 solar spectrum and the atmospheric transmissivity, respectively.  (d) and (e) show typical equilibrium temperatures of the idealized emitter (blue) and absorber (red) with respect to the ambient temperature (green) during hot and cold days. In both plots the nonzero absorptivity and emissivity were chosen as  $a = 0.25$ and $e = 1$, and we modeled the time-dependent total solar irradiance using a gaussian distribution between 6 h and 18 h with peak irradiance of 900 W/m$^2$ at noon. The ambient temperature profile is $T_{\rm amb}(t) =T_{\rm amb}^{\rm avg} + \Delta T_{\rm amb} \sin\left[2\pi(t({\rm h})-11)/24\right]$ with $T_{\rm amb}^{\rm avg}  = 25 ^{o}$C, $\Delta T_{\rm amb}  = 5 ^{o}$C in (d), and $T_{\rm amb}^{\rm avg}  = 5 ^{o}$C, $\Delta T_{\rm amb}  = 10 ^{o}$C in (e).} 
\label{Fig1}
\end{figure}

Let us first consider a simplified model for an ideal mid-IR emitter and a solar absorber to explain the concept of the passive radiative thermostat (Figs. \ref{Fig1}a,b). We assume the former has unit emissivity $e(\lambda,\theta)=1$ within the first atmospheric transparency window $(8\ \mu{\rm m} \leq \lambda \leq 13\ \mu{\rm m})$ and $e(\lambda,\theta)=0$ outside this wavelength range, while the latter has 
constant absorptivity $a(\lambda,\theta) <1$ in the visible and near-IR range $(0.3\ \mu{\rm m} \leq \lambda \leq 2.5\ \mu{\rm m})$ and is zero otherwise (Fig. \ref{Fig1}c). In our toy model, we consider that both  $e(\lambda,\theta)$ and $a(\lambda,\theta)$ are independent of the direction into which radiation is emitted or absorbed from. The net power exchanged between the emitter (absorber) of area $A$ at temperature $T$ and the environment is given by 
\begin{equation}
\label{Pnet}
P_{\rm net}(T)\!=\! P_{\rm rad}(T)\! -\!P_{\rm sun}\! -\!P_{\rm atm}(T_{\rm amb})\! -\!P_{\rm cc}(T, T_{\rm amb}).
\end{equation}
The system emits thermal radiation with a total power given by  
\begin{equation}
\label{Prad}
P_{\rm rad}(T)\!=\! A\int\! d\Omega \cos\theta\! \int_0^{\infty}\! d\lambda I_{\rm BB}(\lambda,T) e(\lambda,\theta) ,
\end{equation}
where  $I_{\rm BB}(\lambda,T) = (2\pi \hbar c^2/\lambda^5)[e^{2\pi \hbar c/\lambda k_{\rm B}T}-1]^{-1}$ is the spectral radiance of a blackbody in thermal equilibrium at temperature $T$, and the solid angle integral is over a hemisphere. The body also absorbs radiation incoming from the sun at a rate 
\begin{equation}
\label{Psun}
P_{\rm sun}\!=\! A\! \cos\theta_{\rm sun}\! \int_0^{\infty}\! d\lambda I_{\rm AM1.5}(\lambda) a(\lambda,\theta_{\rm sun}),
\end{equation}
where $I_{\rm AM1.5}(\lambda)$ is the AM1.5 solar spectral radiance \cite{ASTM2012}, and we assume the planar radiator is facing the sun at an angle $\theta_{\rm sun}$. The body also absorbs heat from the atmosphere at a rate 
\begin{equation}
\label{Patm}
P_{\rm atm}(T_{\rm amb}) \!\!=\!\! A\!\!\int\!\! d\Omega \cos\theta\!\! \int_0^{\infty}\!\! d\lambda I_{\rm BB}(\lambda,T_{\rm amb}) a(\lambda,\theta) e_{\rm atm}(\lambda,\theta), 
\end{equation}
where $e_{\rm atm}(\lambda,\theta) = 1-t(\lambda)^{1/\cos\theta}$ is the angle dependent emissivity of the atmosphere \cite{Granqvist1981} and $t(\lambda)$ corresponds to the atmospheric transmissivity in the zenith direction \cite{Gemini2018}. In principle, the emitter (absorber) can also exchange heat non-radiatively with its surroundings via convection and conduction channels, that we account for through the last term in Eq. (\ref{Pnet}), where $P_{\rm cc}(T,T_{\rm amb}) =  A h (T_{\rm amb}-T)$ and $h$ is the heat coefficient (we assume $h \simeq 7$ W/m$^2$K throughout the paper, which is similar to that reported in 
 \cite{Raman2014}). As the ambient temperature and solar irradiation conditions change during the day, the equilibrium temperature of the body, calculated by enforcing $P_{\rm net}(T_{\rm eq})=0$, will correspondingly vary. Figs. \ref{Fig1}d,e show the equilibrium temperature for both the idealized emitter (blue) and absorber (red) as a function of the time of the day for two simplified ambient temperature profiles $T_{\rm amb}(t)$ (dashed green), corresponding to a typical hot and a cold day, respectively. As expected, the emitter cools below $T_{\rm amb}$ during both day and night, while the absorber heats above $T_{\rm amb}$ during day-time and thermalizes at the ambient temperature during night since $P_{\rm sun}=0$. As is evident from the plots, each of these structures has an intrinsic undesirable feature: the absorber heats well above $T_{\rm amb}$ even in a hot day, while the emitter still cools below $T_{\rm amb}$  in a cold day.  These unwanted temperature swings can be avoided by using materials whose emissivity and absorptivity characteristics are affected by the ambient temperature, and thereby effectively behave as coolers in hot days and as heaters in cold days.

\begin{figure}
\centering
\includegraphics[width=8.6cm]{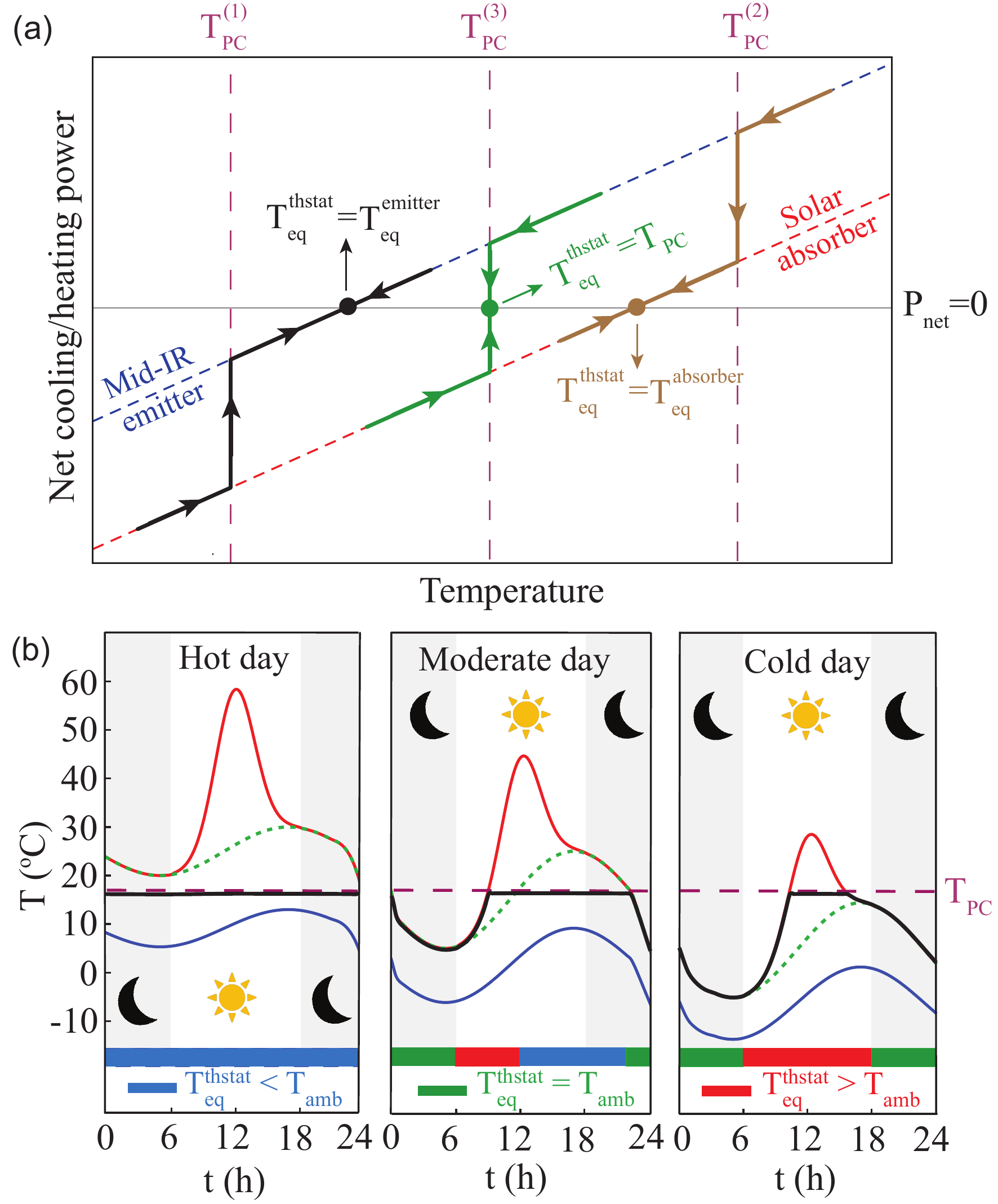}
\caption{Passive radiative thermostat concept. (a) Cartoon representation of the net cooling and heating power versus temperature for three phase-change photonic thermostats  (black, green, and brown curves) with distinct operating phase transition temperatures (vertical lines). The plot also shows the corresponding $P_{\rm net} (T)$ associated to the idealized emitter (blue) and absorber (red) described in Fig. \ref{Fig1}. (b) Comparison between the performance of an idealized emitter (blue), absorber (red), and phase-change photonic thermostat (black) with $T_{\rm PC} = 17 ^o$C (dashed purple) for a hot, moderate, and cold day.
The bottom bars highlight the thermostat's switching between cooling and heating modes during a day depending on the corresponding ambient temperature (dashed green), modeled with parameters
 $T_{\rm amb}^{\rm avg}  = 25 ^{o}$C, $\Delta T_{\rm amb}  = 5 ^{o}$C (left), $T_{\rm amb}^{\rm avg}  = 15 ^{o}$C, $\Delta T_{\rm amb}  = 10 ^{o}$C (center), and $T_{\rm amb}^{\rm avg}  = 5 ^{o}$C, $\Delta T_{\rm amb}  = 10 ^{o}$C (right). All other parameters are the same as in Fig. \ref{Fig1}.} 
\label{Fig2}
\end{figure}

We now introduce the concept of a passive radiative thermostat. Let us consider an idealized body that behaves as our solar absorber for temperatures $T < T_{\rm PC}$, and as our mid-IR emitter for temperatures $T>T_{\rm PC}$, where $T_{\rm PC}$ is the phase transition temperature of the material composing the thermostat. Note that in this case both $P_{\rm sun}$ and $P_{\rm atm}$ will depend on $T$ through the temperature-dependent absorptivity $a_T(\lambda,\theta)$, while $P_{\rm rad}$ will have an additional temperature dependence through the emissivity $e_T(\lambda,\theta)$. In Fig. \ref{Fig2}a we show an schematics of the net exchanged power (solid curves) as a function of the body's temperature for three photonic thermostats with different phase transition thresholds (vertical dashed lines) at an arbitrary time of the day.  In order to most easily  explain the possible dynamical paths towards the steady state of the system in the case of the thermostat, we also show $P_{\rm net}$ for the idealized emitter (dashed blue curve) and absorber (dashed red curve) of Fig. \ref{Fig1} for which no phase transition takes place. The corresponding equilibrium temperatures of the emitter $T^{\rm emitter}_{\rm eq} < T_{\rm amb}$ and  the absorber $T^{\rm absorber}_{\rm eq} \geq T_{\rm amb}$ are shown in the figure. For $T<T_{\rm PC}$, the thermostat curves follow the absorber power curve, while for $T>T_{\rm PC}$ it follows the one for the emitter. 

Let us first consider the case when the phase transition temperature is lower than $T^{\rm emitter}_{\rm eq}$, e.g. $T_{\rm PC}^{(1)}$. When the initial temperature of the body is less than $T_{\rm PC}^{(1)}$, the system will heat up, cross the phase transition, and continue heating until it reaches $T^{\rm thstat}_{\rm eq} = T^{\rm emitter}_{\rm eq}$  (black curve). On the contrary, when the initial temperature is above $T_{\rm PC}^{(1)}$, the body does not undergo any phase transition but still heats up or cools down in order to equilibrate at $T^{\rm emitter}_{\rm eq}$, as indicated by the arrows in the figure. When the phase transition temperature is larger than $T^{\rm absorber}_{\rm eq}$, e.g. $T_{\rm PC}^{(2)}$, the opposite equilibration dynamics takes place (brown curve), and the thermal equilibrium is reached at $T^{\rm thstat}_{\rm eq} = T^{\rm absorber}_{\rm eq}$. Finally, when $T^{\rm emitter}_{\rm eq}<T_{\rm PC}<T^{\rm absorber}_{\rm eq}$, as for the case of $T_{\rm PC}^{(3)}$, the system heats (cools) for initial temperatures smaller (larger) than $T_{\rm PC}^{(3)}$, and locks at the phase transition temperature (green curve). Note that the underlying operation mechanism of the passive radiative thermostat applies for both day- and night-time, as well as it holds for a single $T_{\rm PC}$ and varying $T_{\rm amb}$, since the effect of changing the ambient temperature is simply to shift the curves of $P_{\rm net}(T)$ up or down. 

In general, any of the aforementioned cases can occur during a day depending on the relationship between the phase transition temperature of the thermostat, and the emitter  and absorber equilibrium temperatures. Figs. \ref{Fig2}b-d show the daily equilibrium temperature variation of the idealized emitter (blue), the idealized absorber (red), as well as the idealized passive radiative thermostat (black) with phase transition temperature fixed at $T_{\rm PC}=17 ^o$C for three distinct ambient temperature profiles (dashed green), corresponding to hot, moderate, and cold days. For the chosen parameters, $T^{\rm emitter}_{\rm eq}<T_{\rm PC}<T^{\rm absorber}_{\rm eq}$ (case of $T_{\rm PC}^{(3)}$ above) during the whole 24 hours period in the hot day, resulting in an equilibrium temperature locked at $T^{\rm thstat}_{\rm eq}=T_{\rm PC}<T_{\rm amb}$ and the thermostat working as a radiative cooler all day long. For the moderate day, $T_{\rm PC}>T^{\rm absorber}_{\rm eq}$ (as the $T_{\rm PC}^{(2)}$ case above) for $0<t<9$ h and $22<t<24$ h, so the equilibrium temperature follows the absorber steady state temperature  during those times,  $T^{\rm thstat}_{\rm eq} = T^{\rm absorber}_{\rm eq} \ge T_{\rm amb}$,  with the equality holding at night and the device operating as a radiative heater early in the morning. In the time-range $9<t<22$ h, the phase transition temperature is in-between the absorber and emitter temperature curves, hence $T^{\rm thstat}_{\rm eq}$ is locked at $T_{\rm PC}$, switching  from heating to cooling with respect to the ambient temperature as $T_{\rm amb}$ crosses $T_{\rm PC}$. For the last plot a similar reasoning applies but, in contrast to the case of the moderate day, the locking at $T_{\rm PC}$ results only in day-time heating above the ambient temperature, as desired for a cold day (unfortunately, practical passive heating during night-time is not possible since $P_{\rm sun}=0$). For the chosen parameters in Fig. \ref{Fig2}b, the case of $T_{\rm PC}^{(1)}$ (phase transition temperature below $T^{\rm emitter}_{\rm eq}$) never occurs. Such a case could be achieved, for example, using a lower $T_{\rm PC}$ and would result in $T^{\rm thstat}_{\rm eq}$ following the emitter equilibrium temperature.  

\begin{figure*}
\centering
\includegraphics[width=15cm]{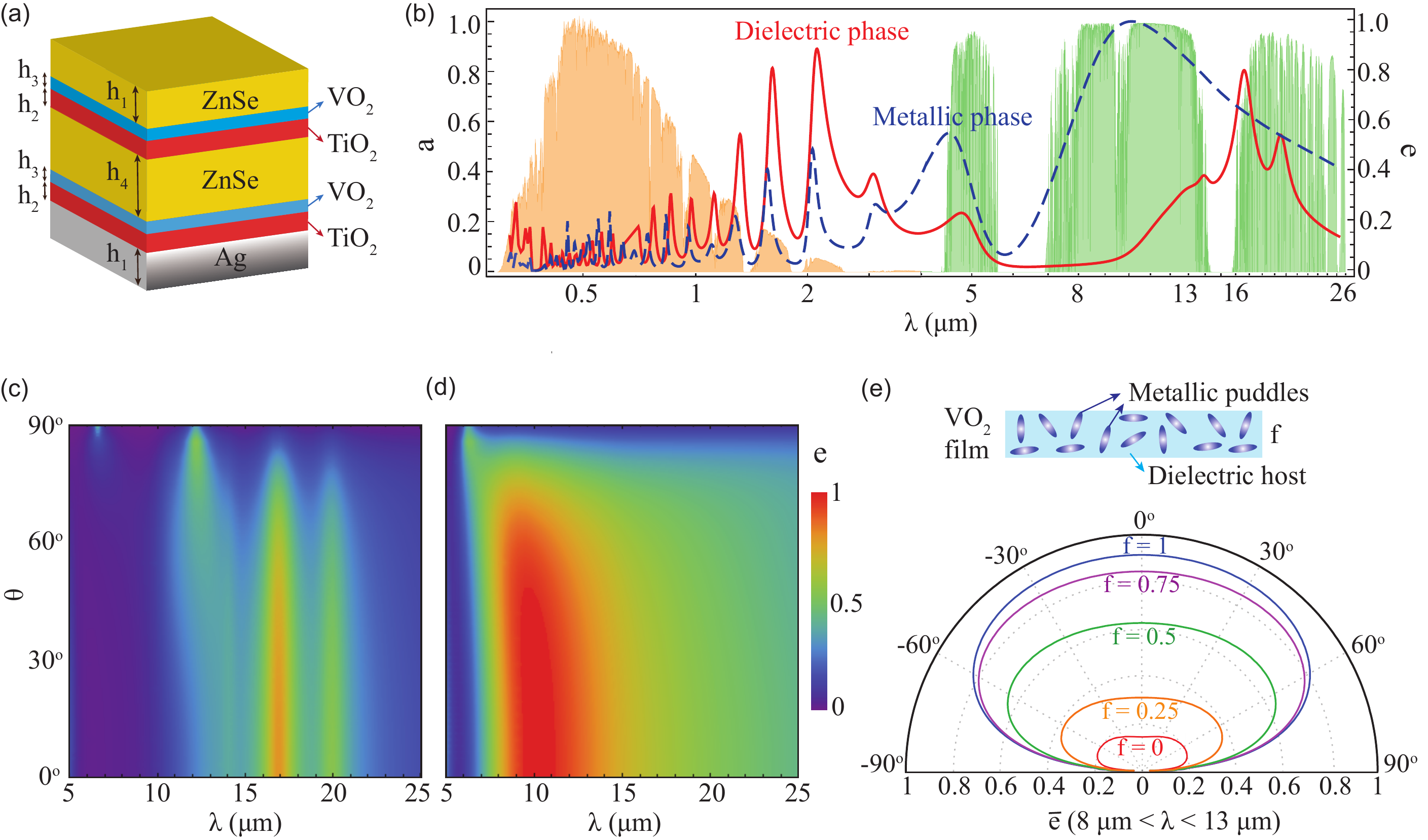}
\caption{Passive radiative thermostat based on a thermochromic VO$_2$ phase-change material. (a) Multilayer photonic nanostructure for passive temperature management with thicknesses $h_1 = 300$ nm, $h_2 = 55$ nm, $h_3 = 30$ nm, and $h_4 = 1\ \mu$m. (b) Calculated absorptivity (= emissivity per Kirchhoff's law) of the designed radiative thermostat versus wavelength for VO$_2$ in the dielectric (red) and metallic (blue) phases for unpolarized radiation. Dependence of the emissivity of our photonic nanostructure on the polar angle at mid-infrared wavelengths for (c) dielectric and (d) metallic phases of VO$_2$. (e) Top: schematics of VO$_2$ metallic puddles (filling factor $f$) in a VO$_2$ dielectric host.  Bottom: angular dependence of the average emissivity in the first atmospheric transparency window $(8\ \mu{\rm m} \leq \lambda \leq 13\ \mu{\rm m})$ across the phase transitions for several filling factors.} 
\label{Fig3}
\end{figure*}

We describe next a practical design to passively achieve temperature moderation using thermochromic phase-change materials. Fig. \ref{Fig3}a shows a VO$_2$-based multilayer photonic nanostructure that presents absorption and emission properties qualitatively similar to those introduced in our toy model above. It is composed of a thick Ag ground plane that prevents transmission in the entire wavelength range of interest, and alternating layers of TiO$_2$, VO$_2$, and ZnSe that provide the necessary emission and absorption features. The physical mechanism behind the optical response of our structure relies on a Fabry-P\'erot cavity specially designed to resonate at mid-IR wavelengths. ZnSe is chosen as the material filling the cavity as it is  almost transparent in the entire  visible to mid-IR wavelength range, allowing for reduced absorption of solar radiation and high emissivity in the first atmospheric transparency window when VO$_2$ is in the metallic phase. Other dielectric substrates (e.g.  Al$_2$O$_3$, SiO$_2$, SiC, Si$_3$N$_4$) previously used to design radiative coolers have phonon resonances  in the mid-IR, resulting in reduced contrast in the optical response of our device below and above the phase transition temperature. In the following we fix $T_{\rm PC}^{\rm VO_2} = 17 ^o$C, consistent with phase-transition temperatures reported in \cite{Muraoka2002, Fan2014, Liu2015} for 30 nm thick VO$_2$ thin films on TiO$_2$. Such $T_{\rm PC}^{\rm VO_2}$ results in operating conditions within the typical ambient temperature variations occurring in Los Alamos, New Mexico. We will assume that the heating and cooling lags associated to VO$_2$ hysteresis occur at close temperatures, and therefore our main conclusions remain unaffected. In the Supporting Information we discuss how material hysteresis may modify the thermostat's equilibrium temperature shown in Fig. \ref{Fig2}. As the insulator-to-metal phase transition of VO$_2$ occurs on a timescale of a few tens of femtoseconds \cite{Jager2017}, we can consider that the system reacts instantaneously to ambient temperature variations that occur at much longer timescales.

In Figs. \ref{Fig3}b-e we show the optical response of our photonic thermostat as a function of different parameters. The absorptivity (= emissivity by Kirchhoff's law) was calculated using the transfer matrix method with $a = 1-r$, where $r$ is the multilayer's reflectivity for unpolarized radiation.  The temperature-dependent refractive index of VO$_2$ was modeled via the Bruggemann effective medium theory for metallic puddles (filling factor $f$) embedded in a dielectric host  \cite{Qazilbash2007,Barker1966, Barker68, Choy199} (see Supplementary Information), while optical data for the other materials were taken from \cite{Ordal1985, Ratzsch2015, Querry1987,RefractiveIndex}. Let us first discuss the emission properties of the thermostat in the mid-IR. Fig. \ref{Fig3}b shows the absorptivity and emissivity at the normal direction for VO$_2$ in its dielectric (red) and metallic (blue) phases.  When VO$_2$ is in the purely dielectric phase ($f = 0$) a strong impedance mismatch at the air-multilayer interface results in decreased absorptivity in the  $8\ \mu{\rm m} \leq \lambda \leq13\ \mu$m range. The emissivity increases near the edge of the first atmospheric transparency window due to enhanced absorption in the TiO$_2$ film, and two peaks emerge at $\sim 17\ \mu$m and  $\sim 20\ \mu$m near the phonon modes of VO$_2$ \cite{Qazilbash2007, Barker1966}. When the VO$_2$ thin films are in the purely metallic phase ($f = 1$) they form a good quality mid-IR Fabry-P\'erot cavity with the fundamental mode resonating at $\lambda \simeq 10\ \mu$m, which leads to near unity emissivity at this wavelength. Note that an additional cooling channel is possible in our system due to the non-zero emission in the atmospheric transparency window  between $20\ \mu{\rm m} \leq \lambda \leq 25\ \mu{\rm m}$. Figs. \ref{Fig3}c and \ref{Fig3}d depict the the system's emissivity dependence over the direction of emission, demonstrating a strong robustness against variations with $\theta$. Fig. \ref{Fig3}e shows polar plots of the average emissivity in the first transparency window, and highlights the quasi-omnidirectional emissivity of the system at various representative stages of the phase transition.  At visible and near-IR wavelengths the absorption in the system takes place in the VO$_2$ thin films, and a small difference can be noticed between the dielectric and metallic phases as a consequence of the weak effect of the phase transition of VO$_2$ at short wavelengths. The solar spectrum weighted average absorptivity of the system at normal incidence is $\sim 15\%$ for the dielectric phase and $\sim 6\%$ for the metallic phase, which is sufficient to empower our radiative thermostat with the desired properties to perform passive temperature regulation.

\begin{figure*}
\centering
\includegraphics[width=16cm]{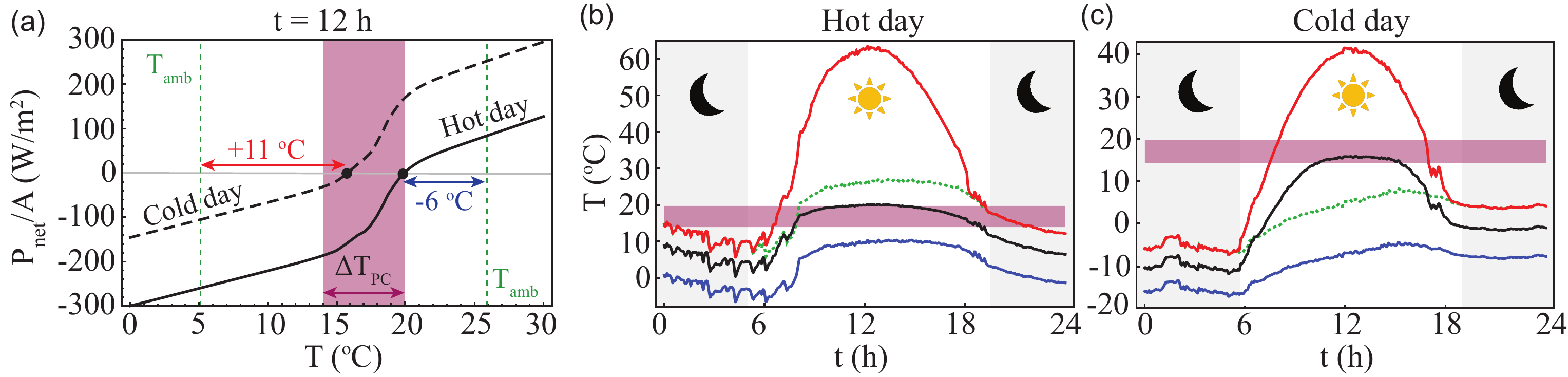}
\caption{Photonic phase-change thermostat performance. (a) Net cooling and heating power of our designed thermostat as a function of temperature on June 23rd, 2017 (hot day, solid curve) and on December 6th, 2017 (cold day, dashed line) in Los Alamos, New Mexico. The vertical dashed lines on the right and left show the respective ambient temperatures at midday, used during the calculations. The black dots indicate the corresponding equilibrium temperatures of the thermostat $T^{\rm thstat}_{\rm eq}(t=12 {\rm h})$, indicating a 
$6 ^o$C cooling ($11 ^o$C heating) with respect to the hot (cold) midday ambient temperature. 
The time-dependent equilibrium temperature of our optimized radiative thermostat (black) for these same days is depicted in (b) and (c) along with the corresponding ambient temperature (green). For comparison, we also show the results for the idealized emitter (blue) and absorber (red) described in Fig. \ref{Fig1}. In all the plots the purplish bands ($\Delta T_{\rm PC}\simeq 6 ^o$C) correspond to the region around the phase transition temperature ($T_{\rm PC} = 17 ^o$C) where $\sim\!90\%$ of the phase-change takes place.} 
\label{Fig4}
\end{figure*}

In Fig. \ref{Fig4} we discuss the performance of our designed passive radiative thermostat under real conditions for operation in winter and summer in Los Alamos. Measured sub-hourly time-dependent ambient temperature and solar irradiation data \cite{Diamond2013, LAdata2017} for two different days with clear sky were used in order to characterize the time-evolution of the equilibrium temperature of the system during both day- and night-time. Fig. \ref{Fig4}a shows the net exchanged power as a function of the thermostat temperature for the cold (dashed curve) and hot (solid curve) days at noon. Unlike the abrupt behavior seen in  Fig. \ref{Fig2}a for the toy model, here both  $P_{\rm net}(T)$  curves feature a smooth transitioning from heating ($P_{\rm net}(T) < 0$) to cooling ($P_{\rm net}(T)>0$) as the temperature crosses the insulator-to-metal phase transition region for realistic VO$_2$ (see Supplementary Information). The day-time equilibrium temperature is slighted shifted with respect to $T_{\rm PC}^{\rm VO_2}$ and thermalization occurs within the range $\Delta T_{\rm PC} = (17 \pm 3) ^o$C, where most of the phase transition takes place. Our passive radiative thermostat provides a remarkable high cooling power (per area) $\sim 100$ W/m$^2$ at temperatures above $T_{\rm PC}^{\rm VO_2}$  during the hot day, and a comparable heating power below the phase transition in the cold day. This results in an equilibrium temperature $6 ^o$C below ($11 ^o$C above) midday ambient temperature in summer (winter). Figs. \ref{Fig4}b,c show the steady state temperature of the thermostat as the environment conditions change. Note that we get a moderation effect on the equilibrium temperature, which remais approximately locked at  $\sim 20 ^o$C ($\sim 15 ^o$C) around noon, regardless of variations in the ambient temperature during the hot (cold) day. In both cases day-time thermalization occurs at temperatures much more comfortable than the undesirable extreme cases achievable by regular radiative absorbers (red) and emitters (blue). In contrast to the idealized case, at night-time we get a net cooling effect arising from the non-zero emissivity in the atmospheric transparency windows even when VO$_2$ is in its dielectric phase; however, the thermostat's temperature is still above the one achieved with the plain radiative emitter (blue curves).  Finally, we comment that locations with distinct variation ranges of the ambient temperature will require different designs of the passive radiative thermostat, with the appropriate phase transition setpoint constrained by the environment's yearlong thermal swings. Given the high tunability of $T_{\rm PC}^{\rm VO_2}$ achieved via various approaches \cite{Wei2009, Cao2009, Driscoll2009, Tan2012,Lee2013, Lopez2002, Lopez2004,Whittaker2009, Yoon2016, Muraoka2002, Fan2014, Liu2015}, it should be straightforward to design a thermostat to operate under climate conditions different from those in Los Alamos.  Provided that $T_{\rm PC}^{\rm VO_2}$ is properly set for a given location, however, all conclusions above should hold. 

In summary, we investigated passive radiative thermostats based on phase-change materials for near room temperature thermal moderation. Such devices moderate temperature swings with respect to plain emitters or absorbers, resulting in both night- and day-time cooling in hot days at a pre-set phase transition temperature, and realizing day-time heating in cold days. We proposed a VO$_2$-based nanophotonic thermostat multilayer design that, at temperatures below the phase transition heats up by absorbing solar radiation, while at temperatures above the phase transition it cools down by emitting electromagnetic energy to outer space through the atmospheric transparency windows.  Additional improvements in the absorptivity and emissivity properties of the thermostat could be achieved  through the introduction of metasurfaces to tailor material resonances at the visible and mid-infrared wavelengths. A possible criteria  to quantify the performance of a given thermostat design is to take the time-average of the absolute value of the difference  between the equilibrium and setpoint temperatures normalized to the thermochromic material typical phase transition band. The smaller the value of this figure of merit the better is the thermostat performance. The passive thermal  moderation mechanism could be used to reduce the impact of material stresses on structures such as bridges undergoing thermal cycles. In addition, given that about $32\%\ (33\%)$ from the total energy consumed in residential (commercial) buildings is used for space heating and $9\%\ (7\%)$ for cooling \cite{Diana2015}, our passive radiative thermostat concept has the potential to largely increase energy savings worldwide.

\begin{acknowledgement}
The authors thank Los Alamos National Laboratory (LANL) Laboratory Directed Research and Development (LDRD) program and the Center for Nonlinear Studies (CNLS) for support. 
\end{acknowledgement}

\begin{suppinfo}
The following files are available free of charge.
\begin{itemize}
  \item Supporting Information. effect of hysteresis on the photonic thermostat performance, VO$_2$ dielectric constant calculation
  \end{itemize}
\end{suppinfo}

\section*{Author Contributions}
WKK conceived the idea, WKK, DD,  and MR developed the passive radiative thermostat concept. WKK and DD wrote the paper with inputs from MR, AA, and SK. 
All authors discussed the results, commented on and revised the manuscript.
\section*{Notes}
The authors declare no competing financial interests.
%

\let\oldsection\section
\renewcommand\section{\clearpage\oldsection}
\section*{Supporting information}

\renewcommand{\thefigure}{S\arabic{figure}}
\setcounter{figure}{0}    
\begin{figure*}[!b]
\centering
\includegraphics[width=12cm]{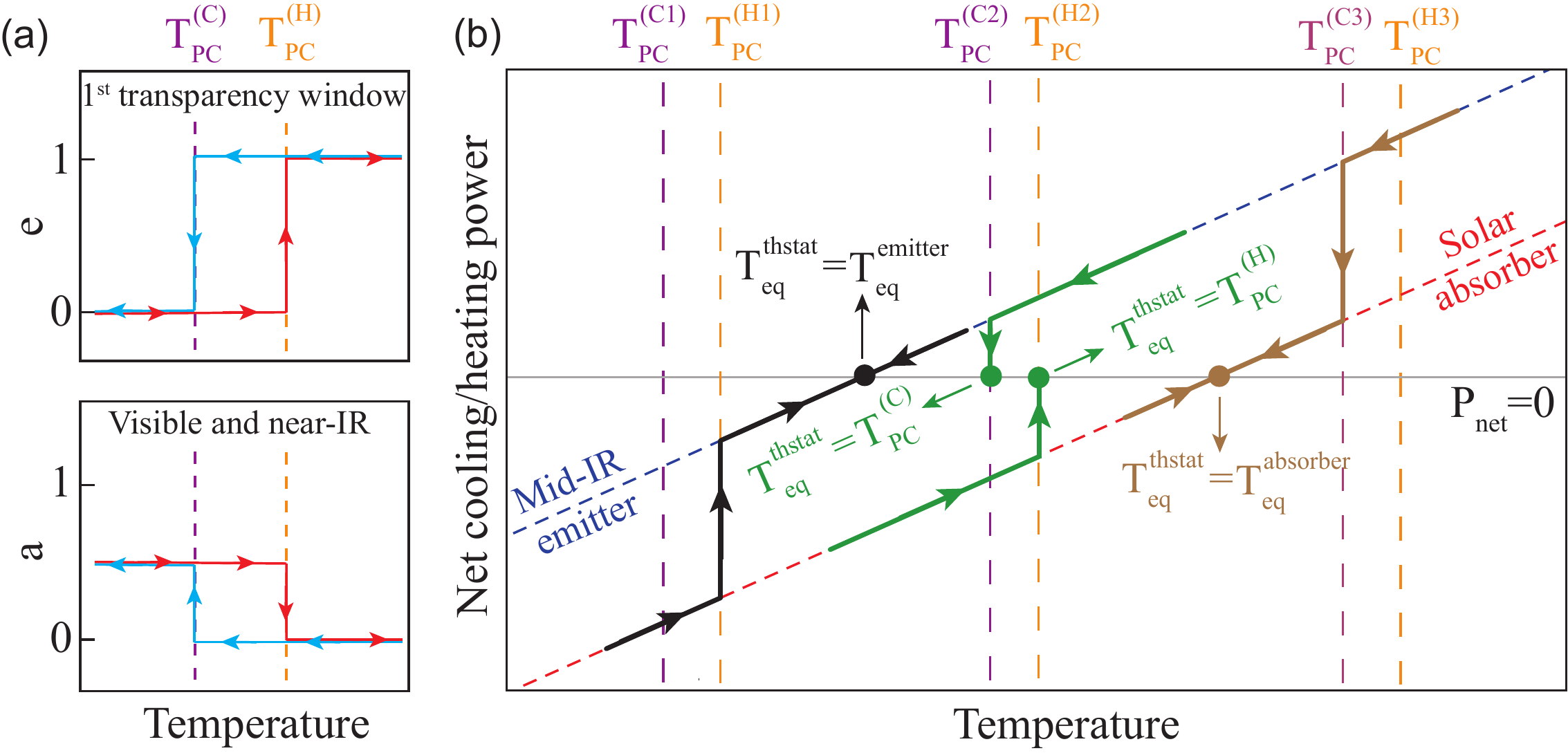}
\caption{(a) Simple model of hysteresis in the thermostat's emissivity (top) and absorptivity (bottom). (b) Cartoon representation of the net cooling and heating power versus temperature for three phase-change photonic thermostats  (black, green, and brown curves) with distinct hysteresis loops. The plot also shows the corresponding $P_{\rm net} (T)$ associated to the idealized emitter (blue) and absorber (red) described in Fig. 1 of the main paper. } 
\label{FigS1}
\end{figure*}

{\bf Effect of hysteresis on the  photonic thermostat performance}  - Let us consider a simplified model for computing effects of hysteresis in a photonic radiative thermostat. We consider a two-step hysteresis loop with transition temperatures $T_{\rm PC}^{\rm H} > T_{\rm PC}^{\rm C}$  during heating and cooling processes, respectively. The thermostat is assumed to behave as the idealized solar absorber introduced in Fig. 1 of the main manuscript for temperatures $T < T_{\rm PC}^{\rm C}$, and as the idealized emitter for  $T > T_{\rm PC}^{\rm H}$. In the temperature range $ T_{\rm PC}^{\rm C} < T < T_{\rm PC}^{\rm H}$ the thermostat presents unity emissivity and zero absorption during cooling, while it shows zero emission and finite absorption when heating. Figure \ref{FigS1}a shows hysteresis cycles for the emissivity and absorptivity in the first transparency window and at visible/near-IR wavelengths, respectively.   In Fig. \ref{FigS1}b we show an schematics of the net exchanged power (solid curves) as a function of the body's temperature for three photonic thermostats with different hysteresis cycles at an arbitrary time of the day. For $T<T_{\rm PC}^{\rm C}$, the thermostat curves follow the absorber power curve, while for $T>T_{\rm PC}^{\rm H}$ it follows the one for the emitter. For temperatures in between, the thermostat will follow the emitter or absorber curves depending on its previous cooling or heating history. There are only four possible equilibrium temperatures for the thermostat, 
$T_{\rm PC}^{\rm C}, T_{\rm PC}^{\rm H}, T_{\rm eq}^{\rm emitter}$, and  $T^{\rm absorber}_{\rm eq}$. When both transition temperatures are below (above) $T^{\rm emitter}_{\rm eq}$ ($T^{\rm absorber}_{\rm eq}$) the thermostats equilibrates at the emitter's (absorber's) equilibrium temperature, regardless of the hysteresis (black and brown curves). When both thresholds are in between $T^{\rm emitter}_{\rm eq}$ and $T^{\rm absorber}_{\rm eq}$ (green curves), the equilibrium temperature is locked either at $T_{\rm PC}^{\rm C}$ or $T_{\rm PC}^{\rm H}$ depending if the system is in a cooling or heating path. The remaining three possible cases not shown in the figure can be analyzed in a similar fashion. For example, for  $T_{\rm PC}^{\rm C}< T^{\rm emitter}_{\rm eq}< T_{\rm PC}^{\rm H}$, $T_{\rm eq}^{\rm thstat} =  T_{\rm PC}^{\rm H}$ during the heating cycle and $T_{\rm eq}^{\rm thstat} =  T^{\rm emitter}_{\rm eq}$ for cooling. 
\begin{figure*}[!t]
\centering
\includegraphics[width=12cm]{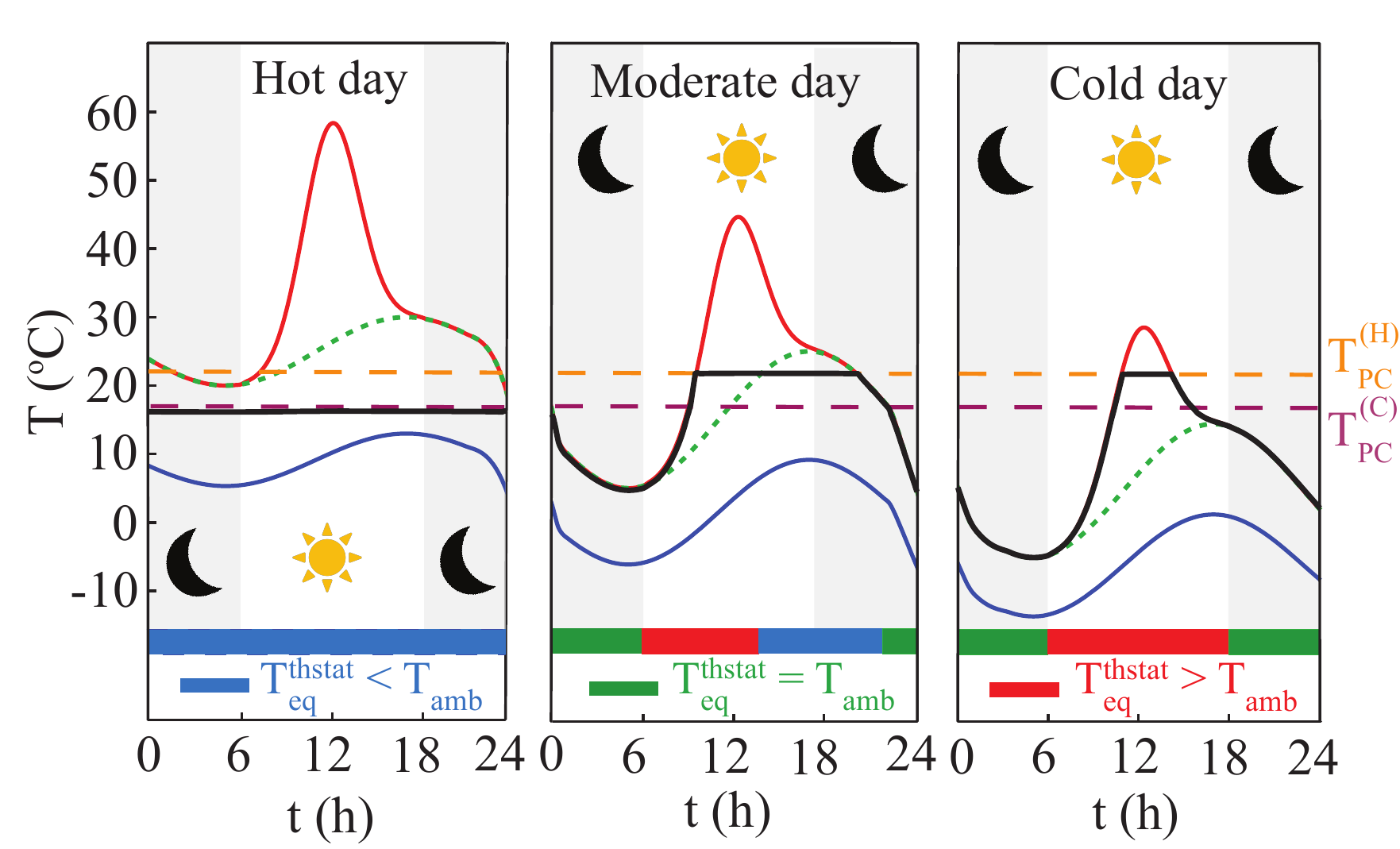}
\caption{(b) Comparison between the performance of an idealized emitter (blue), absorber (red), and phase-change photonic thermostat (black) with $T_{\rm PC}^{\rm C} = 17 ^o$C (dashed purple) and $T_{\rm PC}^{\rm H} = 22 ^o$C (dashed orange) for a hot, moderate, and cold day.
The bottom bars highlight the thermostat's switching between cooling and heating modes during a day depending on the corresponding ambient temperature (dashed green). All parameters are the same as those in Fig. 2 of the main paper.} 
\label{FigS2}
\end{figure*}

In Figure \ref{FigS2} we show the impact of hysteresis on the time-dependent equilibrium temperature for hot, moderate, and cold days (compare to Fig. 2b of the main paper where hysteresis is not taken into account). Assuming that at midnight the system is in a cooling path, we observe temperature locking at $T_{\rm PC}^{\rm C}$ during the entire hot day, and locking at $T_{\rm PC}^{\rm H}$ during daytime for moderate and cold ambient temperatures. During night time there is no difference with respect to the case without hysteresis of the manuscript. Based on these results we conclude that provided both phase transition temperatures are close, the hysteresis does not qualitatively affect the main conclusions of the main paper. We expect that the use of more accurate models of the hysteresis curves in VO$_2$ that include inner loops, such as the discrete Preisach model \cite{Mayergoyz2003_2}, will result in quasi-locking at temperatures between the two extreme temperatures of the main hysteresis loop.

{\bf Effective medium theory and optical properties of VO$_2$} - We model the optical properties of VO$_2$ in an arbitrary stage of the phase transition using the Bruggemann effective medium theory (BEMT) \cite{Choy199_2}. This  is the simplest analytical model that
predicts an insulator-to-metal transition at a critical concentration of metallic puddles in a dielectric host, and it has been successfully employed in previous experimental works to describe the dielectric constant of VO$_2$ \cite{Qazilbash2007_2}. In the BEMT the metallic inclusions are treated as randomly distributed and oriented spheroidal grains with filling factor $f$  and depolarization factor $L$. 
We assume the VO$_2$ effective medium has uniaxial optical response \cite{Barker1966_2, Barker68_2}  described via the dielectric constant tensor $\overline{\varepsilon}_{\rm eff}(\lambda, T) = {\rm diag}[\varepsilon^{\|}_{\rm eff}(\lambda, T), \varepsilon^{\|}_{\rm eff}(\lambda, T), \varepsilon^{\perp}_{\rm eff}(\lambda, T)]$, where the components parallel ($\varepsilon^{\|}_{\rm eff}$) and perpendicular ($\varepsilon^{\perp}_{\rm eff}$)  to the thin film interface satisfy \cite{Choy199_2}
\begin{eqnarray}
(1-f)\! \left\{\!\frac{\varepsilon_{\rm d}^{\|,\perp} - \varepsilon_{\rm eff}^{\|, \perp}}{\varepsilon_{\rm eff}^{\|, \perp} + L_{\|, \perp} \,(\varepsilon_{\rm d}^{\|,\perp}-\varepsilon_{\rm eff}^{\|, \perp})}
+ \frac{4(\varepsilon_{\rm d}^{\|,\perp} - \varepsilon_{\rm eff}^{\|, \perp})}{2\,\varepsilon_{\rm eff}^{\|, \perp}+ (1-L_{\|, \perp})(\varepsilon_{\rm d}^{\|,\perp}-\varepsilon_{\rm eff}^{\|, \perp})}\right\} \nonumber \\
+f\left\{\frac{\varepsilon_{\rm m}- \varepsilon_{\rm eff}^{\|, \perp}}{\varepsilon_{\rm eff}^{\|, \perp}+ L_{\|, \perp} \, (\varepsilon_{\rm m}-\varepsilon_{\rm eff}^{\|, \perp})}
+ \frac{4(\varepsilon_{\rm m} -\varepsilon_{\rm eff}^{\|, \perp})}{2\varepsilon_{\rm eff}^{\|, \perp} + (1-L_{\|, \perp})(\varepsilon_{\rm m}-\varepsilon_{\rm eff}^{\|, \perp})}\right\}
=0, \, 
\label{eq:BEMTVO2} 
\end{eqnarray}
and we have omitted wavelength and temperature dependence for clarity. Here, $\varepsilon_{\rm d}^{\|,\perp}(\lambda)$  and $\varepsilon_{\rm m}(\lambda)$ are the dielectric constants of VO$_2$ in the purely dielectric ($T \ll T_{\rm PC}^{\rm VO_2}, \, f = 0$) and metallic ($T \gg T_{\rm PC}^{\rm VO_2}, \, f = 1$) phases \cite{Barker1966_2, Barker68_2, Qazilbash2007_2}, respectively. Note that in the metallic phase $\varepsilon_{\rm m} = \varepsilon_{\rm m}^{\|} = \varepsilon_{\rm m}^{\perp}$. In our numerical calculations in the main paper we have neglected hysteresis and have modeled the temperature-dependent filling factor via a logistic function  $f(T) = [1+e^{-k(T-{T_{\rm PC}^{\rm VO_2})}}]^{-1}$ where the steepness of the curve is chosen as $k = 1K^{-1}$. The depolarization factor $L_{\|}$ is a function of the concentration $f(T)$ of metallic puddles and was obtained by fitting the experimental data of Ref. \cite{Qazilbash2007}, while $L_{\perp} = 1-2L_{\|}$.

\end{document}